\documentstyle[12pt]{article}

\begin{document}


\def\a{\alpha}
\def\b{\beta}
\def\c{\varepsilon}
\def\d{\delta}
\def\e{\epsilon}
\def\f{\phi}
\def\g{\gamma}
\def\h{\theta}
\def\k{\kappa}
\def\l{\lambda}
\def\m{\mu}
\def\n{\nu}
\def\p{\psi}
\def\q{\partial}
\def\r{\rho}
\def\s{\sigma}
\def\t{\tau}
\def\u{\upsilon}
\def\w{\omega}
\def\x{\xi}
\def\y{\eta}
\def\z{\zeta}
\def\D{\Delta}
\def\G{\Gamma}
\def\L{\Lambda}
\def\F{\Phi}
\def\P{\Psi}
\def\S{\Sigma}

\def\o{\over}
\def\beq{\begin{eqnarray}}
\def\eeq{\end{eqnarray}}
\newcommand{\gsim}{ \mathop{}_{\textstyle \sim}^{\textstyle >} }
\newcommand{\lsim}{ \mathop{}_{\textstyle \sim}^{\textstyle <} }

\def\IJMP{Int.~J.~Mod.~Phys. }
\def\MPL{Mod.~Phys.~Lett. }
\def\NP{Nucl.~Phys. }
\def\PL{Phys.~Lett. }
\def\PR{Phys.~Rev. }
\def\PRL{Phys.~Rev.~Lett. }
\def\PTP{Prog.~Theor.~Phys. }
\def\ZP{Z.~Phys. }


\baselineskip 0.7cm

\begin{titlepage}
\begin{flushright}
UT-860
\\
October, 1999
\end{flushright}

\vskip 1.35cm
\begin{center}
{\large \bf
On Nonlinear Gauge Theory \\ from a Deformation Theory Perspective
}
\vskip 1.2cm
Izawa K.-I.
\vskip 0.4cm

{\it Department of Physics and RESCEU, University of Tokyo,\\
     Tokyo 113-0033, Japan}

\vskip 1.5cm

\abstract{
Nonlinear gauge theory is a gauge theory based on a nonlinear Lie algebra
(finite W algebra) or a Poisson algebra, which yields
a canonical star product for deformation quantization
as a correlator on a disk.
We pursue nontrivial deformation of topological gauge theory
with conjugate scalars in two dimensions.
This leads to two-dimensional nonlinear gauge theory exclusively,
which implies its essential uniqueness.
We also consider a possible generalization to higher dimensions.
}
\end{center}
\end{titlepage}

\setcounter{page}{2}


Nonabelian Chern-Simons gauge theory,
a topological field theory (TFT) of Schwarz type
\cite{Bir},
provides an intriguing framework to deal with
three-dimensional pure gravity on one hand 
\cite{Roc}
and knot invariants on the other
\cite{Wit}.
Two-dimensional
analogue of such a framework is given by nonlinear gauge theory
\cite{Ike,Iza,Sch,Boe},
which is a gauge theory based on a nonlinear Lie algebra
(finite W algebra
\cite{Boe})
or a Poisson algebra
\cite{Str}.
\footnote{More precisely, the structure constants of the algebra
determine the couplings in the theory,
as nonabelian gauge theory is a gauge theory based on a nonabelian
Lie algebra, whose structure constants yield the couplings
of the gauge interaction.}
That is, nonlinear gauge theory
provides a TFT framework to deal with
two-dimensional pure gravity (dilaton gravity
\footnote{The metric field along with the dilaton yields
no local physical degrees of freedom in two dimensions, as is the case for
the metric field in three dimensions.})
on one hand
\cite{Iza,Sch}
and star products on the other
\cite{Kon,Cat}.

Nonlinear gauge theory was originally constructed in part
by inspection
\cite{Ike,Iza}.
In this paper, we consider nontrivial deformation
\cite{Bar}
of abelian BF theory
\cite{Bir}
in two dimensions,
which results in nonlinear gauge theory exclusively.
\footnote{Quadratically nonlinear gauge theory
\cite{Ike}
was obtained in a somewhat related approach by Dayi
\cite{Day}.}
Two-dimensional nonlinear gauge theory is unique in this sense,
as is the case for Chern-Simons gauge theory
in three dimensions
\cite{Bar}.

We first recapitulate a few aspects of nonlinear gauge theory.
The two-dimensional theory is given by an action functional
\cite{Iza}
\footnote{This action is intrinsically two-dimensional
\cite{Ike,Iza},
as the Chern-Simons action is peculiar to three dimensions.}
\beq
 S = \int_\S h^a d \f_a + {1 \o 2} W_{ab}(\f) h^a h^b,
 \label{ACT}
\eeq
where $\f_a$ is a scalar field, $h^a$ is a one-form gauge field
with an internal index $a$
and $d$ denotes an exterior derivative
on a two-dimensional spacetime or worldsheet $\S$.
The structure function $W_{ab}$ is determined by
a nonlinear Lie algebra (finite W algebra
\cite{Boe})
$[T_a, T_b] = W_{ab}(T)$
or a Poisson bracket $[\f_a, \f_b] = W_{ab}(\f)$
of the coordinates $\f_a$ and $\f_b$ on a Poisson manifold $M$,
as was pointed out by Schaller and Strobl
\cite{Str}.
This action is invariant under the nonlinear gauge transformation
\beq
 \d \f_a = W_{ba}\c^b, \quad \d h^a = d \c^a + {\q W_{bc} \o \q \f_a}h^b \c^c
 \label{NGT}
\eeq
due to the Jacobi identity satisfied by $W_{ab}$
\cite{Iza}.

With an ordinary Lie algebra, this theory reduces
to nonabelian BF theory in two dimensions.
When we adopt a nonlinear extension of the two-dimensional
Poincar{\' e} algebra, we obtain the Palatini form of
two-dimensional dilaton gravity
\cite{Iza},
as the Palatini form of three-dimensional gravity is obtained as
Chern-Simons gauge theory of the three-dimensional Poincar{\' e} algebra
\cite{Roc}.

When the Poisson bracket is nondegenerate
to yield a symplectic form $\omega$ on $M$,
we can formally integrate out the fields $h^a$
to obtain an action
\beq
 S = \int_{\q \S} d^{-1} \omega
\eeq
of particles with the phase space $M$ and the vanishing Hamiltonian,
where we adopt a disk as the worldsheet $\S$
and thus $\q \S$ is the boundary circle.
Then the path integral
\beq
 \int \! {\cal D}\f \,  \d_x(\f(\infty)) f(\f(1)) g(\f(0)) \exp {i \o \hbar}S
\eeq
yields an estimation at a point $x \in M$
of the (time-)ordered product of functions
$f$ and $g$ on $M$,
where $\d_x$ is a delta function centered at $x$
and the arguments $0$, $1$ and $\infty$ of $\f$ denote
three points on the circle $\q \S$, placed counterclockwise.

More generally,
a canonical star product given explicitly by Kontsevich
\cite{Kon}
for deformation quantization
\cite{Bay} 
is obtained perturbatively as a correlator
\beq
 f \star g(x) = \int \! {\cal D}\f {\cal D}h \,  \d_x(\f(\infty))
                f(\f(1)) g(\f(0)) \exp {i \o \hbar}S,
\eeq
where $S$ is the action
Eq.(\ref{ACT})
of the nonlinear gauge theory on the disk
$\S$, as was elucidated by Cattaneo and Felder
\cite{Cat}
(we have omitted the ghost part in the above functional integral).

Let us now consider nontrivial deformation of abelian BF theory
in two dimensions
with the aid of the Barnich-Henneaux approach
\cite{Bar}
based on the Batalin-Vilkovisky formalism
\cite{Bat},
where a solution to the master equation provides a gauge transformation
and an invariant action simultaneously.

The free action is given by
\beq
 S_0 = \int \! d^2\x \, \e^{\mu \nu} h_\mu^a \q_\nu \f_a,
\eeq
which is invariant under the gauge transformation
\beq
 \d \f_a = 0, \quad \d h_\mu^a = \q_\mu \c^a.
\eeq
The minimal solution to the classical master equation reads
\beq
 {\bar S} = S_0 + \int \! d^2\x \, h_a^{* \mu} \q_\mu c^a
\eeq
and the BRST symmetry is then given by
\beq
 s = \q_\mu c^a {\q \o \q h_\mu^a}
   + \e^{\mu \nu} \q_\nu \f_a {\q \o \q h^{* \mu}_a}
   + \e^{\mu \nu} \q_\mu h_\nu^a {\q \o \q \f^{*a}}
   - \q_\mu h_a^{* \mu} {\q \o \q c_a^*}.
\eeq
A deformation ${\tilde {\cal L}}$ to the Lagrangian should obey
\cite{Bar}
\beq
 s {\tilde {\cal L}} + d a_1 = 0
\eeq
with its descent equations
\beq
 s a_1 + d a_0 = 0, \quad s a_0 = 0.
\eeq
Thus
\beq
 a_0 = -{1 \o 2}f_{ab}(\f)c^ac^b,
\eeq
where $f_{ab}(\f)$ is antisymmetric.
This implies
\beq
 a_1 = {1 \o 2}{\q f_{ab} \o \q \f_c} h_c^* c^ac^b + f_{ab} h^ac^b,
\eeq
where $h^*_{a \mu}= \e_{\mu \nu} h^{* \nu}_a$,
and leads to
\beq
 {\tilde {\cal L}} = -{1 \o 4}{\q^2 f_{ab} \o \q \f_d \q \f_c}
                      h^*_d h^*_c c^a c^b
 + {\q f_{ab} \o \q \f_c} ({1 \o 2}c_c^* c^a c^b - h^*_c h^a c^b)
 - f_{ab} (\f^{*a} c^b - {1 \o 2}h^a h^b).
\eeq
This deformation is consistent
\cite{Bar}
only when the functions $f_{ab}(\f)$ verify the Jacobi identity.
Then the deformed action
\beq
 {\bar S} + \int {\tilde {\cal L}}
\eeq
satisfies the classical master equation.
For the vanishing antifields, it reduces to the action
Eq.(\ref{ACT})
of nonlinear gauge theory with $f_{ab}(\f) = W_{ab}(\f)$.
Hence we exclusively obtain the nonlinear gauge theory
as the deformation of the free theory
in two dimensions. 

We finally consider a possible generalization to higher dimensions.
Let us try a BF-like Lagrangian
\beq
 {\cal L} = A^{a \mu} D_\mu \f_a + {1 \o 2}B^{\mu \nu}_a R_{\mu \nu}^a,
\eeq
where $B^{\mu \nu}_a$ is an antisymmetric tensor and
\beq
 D_\mu \f_a = \q_\mu \f_a + W_{ab}(\f)h_\mu^b, \quad
 R_{\mu \nu}^a = \q_\mu h_\nu^a - \q_\nu h_\mu^a
            + {\q W_{bc} \o \q \f_a} h_\mu^b h_\nu^c.
\eeq
This turns out to be invariant under the nonlinear gauge transformation
Eq.(\ref{NGT}) with
\beq
 \d A^{a \mu} = A^{b \mu}{\q W_{bc} \o \q \f_a} \c^c
              - B^{\mu \nu}_d {\q^2 W_{bc} \o \q \f_d \q \f_a}h_\nu^b \c^c,
 \quad
 \d B^{\mu \nu}_a = B^{\mu \nu}_c {\q W_{ba} \o \q \f_c} \c^b,
\eeq
since
\cite{Iza}
\beq
 & & \d (D_\mu \f_a) = (D_\mu \f_c){\q W_{ba} \o \q \f_c} \c^b,
\nonumber \\
 & & \d R_{\mu \nu}^a = R_{\mu \nu}^b {\q W_{bc} \o \q \f_a} \c^c
 + \left\{(D_\mu \f_d){\q^2 W_{bc} \o \q \f_d \q \f_a}h_\nu^b \c^c
 - (\mu \leftrightarrow \nu)\right\}.
\eeq
Thus deformation theoretic analysis is also called for
in higher-dimensional cases.

To conclude, we have obtained the nonlinear gauge theory exclusively
as the nontrivial deformation of the free gauge theory
with conjugate scalars in two dimensions.
This may be regarded as a partial result of a systematic search for
topological gauge field theories of Schwarz type: Chern-Simons, BF and
nonlinear gauge theories.
They constitute the simplest class of gauge field theories,
with no local physical degrees of freedom,
whereas they have fertile mathematical contents.
This is not so surprising; quantum field theories
are even expected to describe the whole nature effectively,
which is apparently the most complicated in nature
with huge mathematical structures.

\end{document}